\documentclass[a4paper]{article}

\usepackage{latexsym}

\usepackage{doublespace} 

\def\thebibliography#1{\section*{REFERENCES}
    \addcontentsline{toc}{section}{REFERENCES}
    \list
    {\arabic{enumi}. }{\settowidth\labelwidth{[#1]}\leftmargin\labelwidth
    \advance\leftmargin\labelsep
    \usecounter{enumi}} \def\newblock{\hskip .11em plus .33em minus -.07em}
    \sloppy \sfcode`\.=1000\relax}

\newcommand{\hs}{\hspace{0.6cm}} \newcommand{\sg}{\sigma} \newcommand{\Sg}{\Sigma}
\newcommand{\La}{\Lambda} \newcommand{\pa}{{\hat P}} \newcommand{\te}{{\hat T}}
\newcommand{\ca}{{\hat C}} 
\newcommand{\dchi}{\delta\chi} \newcommand{\dpi}{\delta \pi}
\newcommand{\defi}{\stackrel{Df}{=}} \newcommand{\den}{\stackrel{d}{=}}
\newcommand{\rep}{\stackrel{\wedge}{=}}

\newcommand{\lista}[2]{\newcounter{#1}\begin{list}
{$\bf #2_{\arabic{#1}}$}{\usecounter{#1}}}

\newenvironment{proof}{ \footnotesize {\bf Proof:}}

\begin{document}
\section*{Steps towards the axiomatic foundations of the relativistic
  quantum field theory: Spin-statistics, commutation relations and CPT 
theorems\\[6pt]}

\begin{quote}
{\bf Gabriel D. Puccini} \footnote{Departamento de F\'\i{}sica, Universidad
Nacional de La Plata, Argentina. Present address: Instituto de Neurociencias, 
Universidad Miguel Hern\'andez, Apartado 18, 03550 San Juan de Alicante, Spain.}  
{\bf 
and H\'ector Vucetich} \footnote{Observatorio Astron\'omico, Universidad
Nacional de La Plata, Paseo del Bosque S./N., (1900) La Plata, Argentina.}\\[1cm]
\end{quote}
%
%
%
%
\begin{quote} 
A realistic physical axiomatic approach of the relativistic quantum
field theory is presented. Following the action principle of
Schwinger, a covariant and general formulation is obtained. The
correspondence principle is not invoked and the commutation relations
are not postulated but deduced. The most important theorems such as
spin-statistics, and CPT are proved. The theory is constructed form
the notion of {\em basic field} and {\em system of basic fields}. In
comparison with others formulations, in our realistic approach fields
are regarded as real things with symmetry properties. Finally, the
general structure is contrasted with other formulations.\\
\end{quote} 
%
%
%
\noindent{KEYWORDS: Schwinger principle, physical axiomatics, CPT theorem, 
spin-statistics theorem, quantized fields}
%
%
%
\section*{1. INTRODUCTION}

The most effective way of systematizing and elucidating a body of
ideas, enhancing clarity and rigor, is by axiomatizing a theory.
Although it has been proved to be very fruitful in mathematics, it
has rarely been tried in physics. This is due, in part, to that a
physical theory presents an additional difficulty, because it
include a mathematical formalism but it is more than this. This
something more is the  physical meaning. And the way of attaching
a physical meaning to a formalism has been a very elusive problem.
In general, it proceeds informally by the use of analogies and
heuristic clues. But many wrong interpretations originate in an
informal analysis of the structure of the theory. Therefore, we
think that the assignation of meaning must be formal in such a way
that all the presuppositions and interpretation rules be explicit.
Following this idea, our physical axiomatics consists not only of a
logical organization of a theory (as a mathematical systems of
axioms), but also of an adequate characterization of the physical
meaning of the symbolism and an examination of the metatheoretical
aspects. As a consequence, our axiomatic approach has a number of
important advantages: (a) It avoid mistakes in the interpretation
of the theory because the assignation of meaning is made by means
of semantical axioms. (b) It avoid the intrusion of elements that
are alien to physics (e.g. observers) because all the
presuppositions of the theory are made explicit from the
beginning. (c) The physical referent class of the theory (i.e. the
set of physical objects it describes) can be identified from the
primitive basis and the axiom systems. (d) It allows an formal
analysis of the structure of the theory, identifying key concepts
and hypothesis, and facilitating the control of the derivations of
theorems. Armed with this approach, in this paper we present an
axiomatic formulation of the relativistic quantum field theory
(RQFT) and we analyze its general structure.

Before axiomatizing a field theory, it must be clearly stated if the concept of 
field
is justified. This is one of the most disputed questions about foundations of any 
RQFT
and, based on different conceptions, a number of formulations have been developed. 
The
most of these, are oriented either to eliminate the notion of `field' because fields
are unobservables \cite{Wi56, St64, Ha62, Wh37, He43} or to assign them only a
computational role \cite{We64, We96}. Here, differing from the bulk of the
formulations, we propose a framework in which `quantum fields' are the {\em prima
matter} of the theory. That is, we assume a realistic ontology \cite{Bu77, Bu79} in
which fields are regarded as quantum entities (i.e. as real things) with a number of
properties (association, transformation properties, mutual action or interaction,
etc.) and we shall construct the theory from the notion of {\em basic field} and 
{\em
system of basic fields}.\footnote{The ontological concepts presupposed by any field
theory were presented in the Ontological Background in ref.\cite{Pu01}} From this
point of view, the real existence of `fields' is an important assumption about the
basic structure of matter, in terms of which is intended to explain the observed
properties of particles. Moreover, we base our approach on the action principle of
Schwinger \cite{Sc51,Sc53}. This principle is stated as a variational equation for 
the
action integral operator (the spacetime integral of the Lagrange function-operator)
which is an infinitesimal alteration of the transformation function (the 
infinitesimal
generator of unitary transformations). Thus, the action principle of Schwinger 
appear
as a differential version of the integral formulation of quantum mechanics given by
Feynman. As a result of this realistic axiomatic formulation, in this paper we 
deduce
the most important theorems such as spin-statistics, CPT and commutation relations.

Our formulation has the following advantages
when compared with others: (a) It provide a abstract (in
the sense that it does not dependent of any particular
representation of the field operators) and covariant formalism
that does not use of the so called ``correspondence principle". We
think that correspondence principles must be avoided into the
construction of physical theories because it is not a constitutive
principle but an heuristic principle which is useful only as a
conceptual test for compatibility of a theory with less refined
theories. (b) The commutation relations between Bose and Fermi
field operators are not postulated but deduced. (c) The
spin-statistics relation is probed with great generality from the
property of invariance under time reversal. This proof,
differently from the original given by Pauli \cite{Pa40}, follows
from a direct argument either for integer spin or half-integer
spin and in an independent way of the causality requirement, as is
the case of the proof given by Weinberg \cite{We64,We96}. (d) The
CPT theorem is proved from the general form postulated for the
dynamical Lagrangian, and the mathematical properties of this
formulation. Moreover, our proof follows avoiding to analyze the
transformation properties for each kind of field, as has been
proved by L\"uders \cite{Lu57}. (e) In addition, other important
results such as the field equations, the expression for the
generators and the crossing symmetry theorem can also be deduced
\cite{Pu02}. However, it should be stressed that several
mathematical and physical problems will not be addressed in this
paper. Among them, let us mention the mathematical structure of
the distribution-valued field operators and Green functions, the
role of the renormalization group and its physical interpretation,
the structure of gauge symmetries and the associated Ward
identities. Some of these problems will be dealt, hopefully, in
forthcoming papers.

The article is organized as follows. In the second section, we
present the physical axiomatics of RQFT. In the third section, we
deduce the spin-statistics theorem. In the fourth section we
obtain the commutation relations of field operators as a theorem.
In the fifth, we deduce the fundamental theorem that must be
satisfied by any RQFT: the CPT theorem. We give a simple example
of this formulation in the section six. Finally, in the last
section we compare our presentation with others, and we discuss
our results.
%
%
%
\section*{2. PHYSICAL AXIOMATICS OF RQFT}
In this section we shall exhibit the axiomatic structure of RQFT.
Firstly we shall list the set of ideas that the theory takes for
granted. The {\em formal background} consists of all the logical
and mathematical ideas it employs, and the {\em material
background} consists of all the generic and specific physical
theories it presupposes. As we shall see, RQFT presupposes no
specific theory and for this reason it is called a {\em
fundamental theory}.

\subsection*{2.1  FORMAL BACKGROUND}
\lista{Pr}{P}
\item Bivalent logic.
\item Formal semantics \cite{BungeT1,BungeT2}.
\item Mathematical analysis with its presuppositions and theory of generalized
functions \cite{Ge64}.
\item Group theory.
\end{list}

\subsection*{2.2  MATERIAL BACKGROUND}
\lista{Prd}{P} \setcounter{Prd}{4}
\item Protophysics \cite{Bu67} (i.e. Physical probabilities, Chronology,
Physical geometry,\ldots).
\end{list}

\subsection*{2.3  PRIMITIVE BASIS}
The conceptual space of the theory is generated by the basis {\bf
B} of primitive (or undefined) concepts, where:
$$
{\bf B}={\langle M^4, \overline{\Sigma}, \Sigma, {\cal H}_E, {\cal
P},{\cal A},
 \hbar, c, {\cal F},
{\bf P_F} \rangle}\;\;.
$$

The elements of this basis will be characterized both formally and
semantically by the axiomatics of the theory and the derived
theorems. According to their status in the theory, the axioms will
be divided in three classes: mathematical {\bf[M]}, physical {\bf
[P]} and semantical {\bf [S]}. \footnote{The semantical symbol
$\defi$ will be used for `definition', $\rep$ for
 `representation', and $\den$ for the relation of `denotation' (See ref.
 \cite{BungeT1} for details.)}

\subsection*{2.4  AXIOMATIC BASIS}
\hs RQFT is a finite-axiomatizable theory, whose axiomatic basis
is

\[
{\cal B}_{A}(RQFT)=\bigwedge^{70}_{i=1} {\bf A_{i}}
\]

where the index $i$ runs on the axioms.

\subsection*{2.5  DEFINITIONS}
\lista{Df}{D}
\item $K \defi $ set of inertial reference systems.
\item eiv $\hat A \defi$ eigenvalues of $\hat A$.
\item $ \langle \Phi_1 | \Phi_2 \rangle \defi $ scalar product of
vectors $\Phi_1$ and
 $\Phi_2$.
\item An interval $d\tau^2=dx^\mu dx_\mu$ between two
points in spacetime will be called {\em spacelike} if $d\tau^2 <
0$, {\em timelike} if $d\tau^2 > 0$ and {\em lightlike} if
$d\tau^2=0$.
\item A continuous set of physically independent
(i.e. spacelike ) points forms a spacelike surface. Formally:
$(\forall k)_K, s_\mu = \{x_\mu \in R^4 /dx^\mu dx_\mu < 0
\}\;\;.$
\end{list}

\subsection*{2.6  AXIOMS}

\subsubsection*{GROUP I: SPACETIME}

\lista{Ax}{A} \label{espa}
\item {\bf [M]} $M^4 \equiv $ Minkowski four-dimensional space.
\item {\bf [S]} $M^4 \rep $ physical spacetime.
\item {\bf [S]} $(\forall p)_{M^4} (\forall k)_K (\exists x_\mu)_{R^4} (x_\mu
\rep p $ and $ x_\mu \den coordinates).$
\item {\bf [S]} $(\forall k)_K (\forall x_\mu)_{R^4} ( x_\mu \rep $ a
  spacetime point
referred to the inertial reference system $k \in K$. The component
$x_0 \den $ an instant of time and the components $x_i \den $ a
space position $)$.
\end{list}

\subsubsection*{GROUP II: F-SYSTEMS}

\lista{Ax4}{A} \setcounter{Ax4}{4} \label{sist}
\item {\bf [M]} $\Sigma, \overline \Sigma $: nonempty numerable sets.
\item {\bf [S]} $(\forall{\sg_i})_{\Sigma} (\sg_i \den \mbox{a basic field}).$
\item {\bf [S]} $(\forall{\sg})_{\Sigma} ({\cal C}(\sg)=\{\sg_1,...,\sg_N \})
(\sg \den \mbox{f-system}).$ \footnote{${\cal C}(\sg)$ denotes the
composition of $\sg$ (see {\bf Section 2} of ref.\cite{Pu01})}
\item {\bf [S]} $(\forall{\overline \sg})_{\overline \Sigma} (\overline \sg
 \den$ environment of some f-system$)$. In particular, $(\overline \sg_o \den$
 the empty environment $).$
\item {\bf [S]} $(\forall {\sg},{\overline \sg_o})_{\Sigma \times \overline
\Sigma } (\langle \sg, \overline \sg_o \rangle \den \mbox{a closed
f-system}). $ \footnote{We restrict
 here to closed systems and we consider that $\sg$ denotes $\langle \sg, \sg_0
 \rangle$.}
\item {\bf [P]} $(\exists K)((K \subset \overline \Sigma)  \wedge (\mbox{ the
configuration of } k \in K \mbox{is independent of time})).$
\end{list}

\subsubsection*{GROUP III: STATES}

\lista{Ax10}{A} \setcounter{Ax10}{10} 
\item {\bf [M]}\label{espest} $(\forall \sg)_{\Sigma} (\exists {\cal
 H}_E=\langle {\cal
L,H,L'}\rangle \equiv \mbox{rigged Hilbert space}). $
\item {\bf [P]} There exists a one-to-one correspondence between
states of $\sg \in \Sigma $ and rays ${\cal R}_{\sg} \subset {\cal
H}$.
\item {\bf [M]} $(\forall \sg)_{\Sigma} ({\cal C}(\sg)=\{\sg_1,...,\sg_N \}
\Rightarrow {\cal H}_E = \otimes^{N}_{i=1} {\cal H}_{E_i} ).$
\item {\bf [S]} $|\Phi(\sg,k)\rangle \in {\cal R}_{\sg} \den $ basis
vector that is the representative of the ray ${\cal R}_{\sg}$ of the
f-system $\sg$ with respect to $k \in K$.
 \footnote{In order to avoid unnecessary complexity in notation
 we are not going to explicit the dependence of the state on the
 system and on the reference system.}
\item {\bf [P]} $(\exists |0\rangle)_{{\cal R}_\sg} (|0\rangle \den $
normalized vacuum state).
\end{list}

\subsubsection*{GROUP IV: OPERATORS AND PHYSICAL QUANTITIES}

\lista{Ax15}{A} \setcounter{Ax15}{15} \label{propq}
\item {\bf [M]} ${\cal P}\equiv \mbox{nonempty family of applications over }
\Sigma$.
\item {\bf [M]} ${\cal A} \equiv \mbox{ ring of operators over } {\cal H}_E$.
\item {\bf [P]} $(\forall P)_{\cal P} (\exists \sg)_\Sigma (P \in P(\sg))$.
\item {\bf [P]} $(\forall P)_{\cal P} (\exists {\hat A})_{\cal A}({\hat A} \rep
P).$
\item {\bf [P]} ${(\forall \sg)}_{\Sigma} {(\forall {\hat A})}_{\cal A}
{(\forall a)}_{\Re} (eiv  {\hat A}=a \wedge {\hat A} \rep P
\Rightarrow a$ is the sole value that $P$ takes on
 $\sg )$.
\item {\bf [M]} $\hbar = c = 1$.
\item {\bf [M]} (Linearity and Hermiticity) ${(\forall \sg)}_{\Sigma} (\forall
 {\hat A})_{\cal A} \forall P)_{\cal P} ({\hat A} \rep P \wedge |\Phi_1 \rangle,
|\Phi _2 \rangle \in {\cal H}_E \Rightarrow $

\begin{enumerate}
\item $ {\hat A} [\lambda_1 |\Phi_1 \rangle + \lambda_2 |\Phi_2 \rangle ]=
\lambda_1  {\hat A} |\Phi _1 \rangle + \lambda_2  {\hat A} |\Phi_2
\rangle, \mbox{  with } \lambda_1, \lambda_2 \in {\cal C}$,
\item $ {\hat A}^{\dag}= {\hat A}).$
\end{enumerate}
\item {\bf [P]} (Probability Densities) $(\forall \sg)_{\Sigma} (\forall  {\hat
 A})_{\cal A} (\forall P)_{\cal P} (\forall |a \rangle)_{{\cal H}_E} (\forall
|\Phi \rangle )_{{\cal H}_E} ( {\hat A} \rep P \wedge  {\hat A}|a
 \rangle=a |a \rangle \Rightarrow $ the probability density $\langle \Phi |a \rangle
\langle a | \Phi  \rangle $ corresponds to the property $ P $ of
the f-system $ \sg )$.
\item {\bf [P]} (Unitary Operators) $(\forall \sg)_{\Sigma } (\forall {\hat
 A})_{\cal A} (\forall P)_{\cal P} (\forall  {\hat U})(( {\hat A} \rep P) \wedge
({\hat U} $
is an operator on $ {\cal H}_E ) \wedge ({\hat U}^{\dag} = {\hat
U}^{-1}) \Rightarrow  {\hat U} {\hat A} {\hat U}^{-1} \rep P)$.
\end{list}

\subsubsection*{GROUP V: QUANTUM FIELDS AND FIELD OPERATORS}

\lista{Ax24}{A} \setcounter{Ax24}{24}
\item {\bf [M]} ${\cal F}\subset {\cal A} \equiv $ nonempty set of
 differentiable operators over ${\cal H}_E$.
\item {\bf [M]} $(\forall \sg_i)_{\Sigma} (\exists {\hat
\chi}_A)_{\cal F} ( {\hat \chi}_A \equiv ( \hat \chi_{A_1}, \cdots,
\hat \chi_{A_k},\cdots) )$.
\item {\bf [S]} $(\forall \sg_i)_{\Sigma} (\hat \chi_{A}\den $ field
operator in the $A$ representation associated with the basic field
$\sg_i ). $
\item {\bf [S]} $(\forall \sg_i)_{\Sigma} (\forall x)_{M^4} ( {\hat
\chi}_A (x) \rep $ the amplitude of the basic field $\sg_i $ at $x )$.
\item {\bf [M]} \label{field} $(\forall{\sg})_{\Sigma} (\exists {\hat
 \chi})_{\cal F} (\hat \chi \equiv (\hat \chi^1,...,\hat
 \chi^l,...)).$%
\footnote{Here
the index $l$ runs over all the different representations. Usually the
 components
$\hat \chi^{l'}$ of each field operator $\hat \chi^l$ will be mutually
 dependent.}
\item {\bf [S]} $ {\hat \chi} \den $ general field operator associated
with the f-system $\sg $.
\item {\bf [M]} \label{hermit} ${\hat \chi}^l= {{\hat \chi}^{l
\dag}}$.
\end{list}

\subsubsection*{GROUP VI: POINCAR\'E GROUP}

\lista{Ax31}{A} \setcounter{Ax31}{31} \label{poinc}
\item {\bf [S]} ${\bf P_F} \den$ Full Poincar\'e group. \footnote{All
 elements of the Poincar\'e group, together with space inversion
 $I_s$, time reversal $I_t$ and their products is called the Full
 Poincar\'e group.}
\item {\bf [S]} ${\bf P_+} \den$ Proper Inhomogeneous Orthochronous
Lorentz Group or Poincar\'e Group.
\item {\bf [S]} ${\bf L_+} \den$ Proper Homogeneous Orthochronous
Lorentz Group or Lorentz Group.
\item {\bf [S]} $(\forall g)_{\bf P_+} (g =(\La, b) \den$ a generic
element of the Poncar\'e Group, where $ b \den $ a 4-dimensional
translation and $ \La \den$ a Homogeneous Lorentz transformation $)$.
\item {\bf [M]} The structure of Lie algebra of the Poincar\'e group
is generated by the operators $\{ {{\hat H},{\hat P}_i, {\hat K}_i,
{\hat J}_i} \} \subset {\cal A}$.
\item {\bf [S]} ${\hat H} \den $ the time translations generator.
\item {\bf [S]} $(\forall \sg)_{\Sigma} ( eiv {\hat H} \den E \rep $
the energy of $\sg)$.
\item {\bf [S]} ${\hat P}_i \den $ the spatial translations generator .
\item {\bf [S]} $(\forall \sg)_{\Sigma} ( eiv {\hat P}_i \den p_i \rep
$ the i-th component of linear momentum of $\sg) $.
\item {\bf [S]} ${\hat K}_i \den $ the generator of the pure
transformations of Lorentz.
\item {\bf [S]} ${\hat J}_i \den $ the generator of spatial rotations.
\item {\bf [S]} $(\forall \sg)_{\Sigma} ( eiv {\hat J}_i \den j_i \rep
$ the i-th component of angular momentum of $\sg ) $.
\item {\bf [P]} The vacuum state $|0 \rangle$ is the state that is
invariant under Poincar\'e transformations (up to a possible phase
factor accounting for a constant energy).
\end{list}

\subsubsection*{GROUP VII: CONTINUOUS TRANSFORMATIONS: POINCAR\`E}

\lista{Ax44}{A} \setcounter{Ax44}{44} \label{trafield}
\item {\bf [M]} Each A-component of the field operators associated to basic
fields of spin $j$, denoted by {${\hat \chi^{j \lambda}_{AB}}$},
transforms under an arbitrary Poincar\'e transformation as:
$$ \hat U(\La,b) \hat \chi^{j \lambda}_{AB} (x)\hat U^{-1}(\La,b)=
{D^{(j)}_{AB}(\La^{-1})^\lambda}_{\lambda'} \hat \chi^{j
\lambda'}_{AB} (\La x + b),
$$
where the matrix $(2j+1)$-dimensional
${D^{(j)}_{AB}(\La)^\lambda}_{\lambda'}$
 is some irreducible representation
$(A,B)$ of $\La$.\footnote{The representation labelled by $(A,B)$
has a dimensionality $(2A+1)(2B+1)$ where $A$ and $B$ assume
values integer and/or half integer.}
\end{list}

\subsubsection*{GROUP VIII: DISCRETE TRANSFORMATIONS: SPACE INVERSION}

\lista{Ax45}{A} \setcounter{Ax45}{45}
\item {\bf [S]} ${(\forall I_s)}_{\bf P_F} (I_s \den $ space inversion,
represented in the Hilbert space by the unitary operator $\hat P
\den \hat U(I_s) ).$
\item {\bf [M]} \label{spacinv} The field operators associated to
 basic fields transform under spatial inversion as follows:
$$
\pa \hat \chi_{\stackrel{\leftrightarrow}{AB}} (x) \pa^{-1}=
D^{-1}(I_s) \hat \chi_{\stackrel{\leftrightarrow}{AB}}(I_s x) =
\nu_s \chi_{\stackrel{\leftrightarrow}{BA}}(I_s x)
$$
with $\nu_s$ an arbitrary c-number of modulus one, i.e. a phase
factor. Moreover,
$$
\hat \chi_{\stackrel{\leftrightarrow}{AB}}\den \pmatrix{\hat
\chi_{AB}\cr \hat \chi_{BA}}
\mbox{ with }{\stackrel{\leftrightarrow}{AB}}\den
(A,B) \oplus (B,A) \mbox{ for } A  \ne B
$$
and
$$
\hat \chi_{\stackrel{\leftrightarrow}{AB}}\den \hat \chi_{AA}
\;\;\;\;\mbox{with}\;\;\;\;{\stackrel{\leftrightarrow}{AB}}\den
(A,A) \;\;\;\;\mbox{for}\;\;\; A = B.
$$
\end{list}

\subsubsection*{GROUP IX: CHARGE OPERATOR}

\lista{Ax47}{A} \setcounter{Ax47}{47}
\item {\bf [M]} $(\forall \sg_i)_{\Sg} (\exists \hat Q)_{\cal A} (\hat
 Q \mbox{ has a discrete spectrum of real eigenvalues } ).$
\item {\bf [M]} \label{carga} $(\forall \sg_i)_{\Sg} ([\hat Q, \hat
 H]= [\hat Q, \hat P_i]= [\hat Q, \hat K_i]= [\hat Q, \hat J_i]=0 ).$
\item {\bf [S]} $\hat Q \den $ the generator of gauge transformations
of the first kind.
\item {\bf [S]} $(\forall \sg_i)_{\Sg} (eiv \hat Q \den q \rep $ the
charge of the $\sg_i )$.
\item {\bf [P]} The vacuum state $|0 \rangle$ is the state that is
invariant under gauge transformations of the first kind.
\end{list}

\subsubsection*{GROUP X: DISCRETE TRANSFORMATIONS: CHARGE CONJUGATION }

\lista{Ax52}{A} \setcounter{Ax52}{52}
\item {\bf [S]} $I_c \den $ charge conjugation, represented in the
Hilbert space by the unitary operator $\hat C \den \hat U(I_c)$.
\item {\bf [M]} \label{charcon} The field operators associated to
 basic fields transform under charge conjugation as,
$$
\ca {{\hat \chi}_{{\stackrel{\leftrightarrow}{AB}}_{12}}}(x)
\ca^{-1}= D^{-1} (I_c) {{\hat
\chi}_{{\stackrel{\leftrightarrow}{AB}}_{12}}} (x) = \nu_c {{\hat
\chi}_{{\stackrel{\leftrightarrow}{AB}}_{1(-2)}}} (x)
$$
with $\nu_c$ a phase factor. Moreover, for
${\stackrel{\leftrightarrow}{AB_1}} \ne
{\stackrel{\leftrightarrow}{AB_2}}$ :
$$
{\hat \chi_{{\stackrel{\leftrightarrow}{AB}}_{12}}} \den
{\pmatrix{ {\hat\chi_{\stackrel{\leftrightarrow}{AB_1}}} \cr
{\hat \chi_{\stackrel{\leftrightarrow}{AB_2}}} }}
\;\;\;\;\mbox{and}\;\;\;\;{\hat
\chi_{{\stackrel{\leftrightarrow}{AB}}_{1(-2)}}} \den {\pmatrix{
{\hat\chi_{\stackrel{\leftrightarrow}{AB_1}}} \cr  -{\hat
\chi_{\stackrel{\leftrightarrow}{AB_2}}} }}
\;\;\;\;\mbox{with}\;\;\;\;
{{\stackrel{\leftrightarrow}{AB}}_{12}}\den
{\stackrel{\leftrightarrow}{AB_1}} \oplus
{\stackrel{\leftrightarrow}{AB_2}}
$$
and
$$
{\hat \chi_{{\stackrel{\leftrightarrow}{AB}}_{12}}} \den {\hat
\chi_{\stackrel{\leftrightarrow}{AB}}} \;\;\;\;\mbox{with}\;\;\;\;
{{\stackrel{\leftrightarrow}{AB}}_{12}}\den
{\stackrel{\leftrightarrow}{AB}}\;\;\;\;\mbox{ for
}\;\;\;{\stackrel{\leftrightarrow}{AB_1}} =
{\stackrel{\leftrightarrow}{AB_2}}.
$$
\end{list}

\subsubsection*{GROUP XI: CONTINUOUS TRANSFORMATIONS: GAUGE}

\lista{Ax54}{A} \setcounter{Ax54}{54}
\item {\bf [M]} \label{gauge} The field operators associated to basic fields
 transform under a gauge transformation of the first kind as:
$$
{\hat U}(\tau) \hat \chi^l (x) {\hat U}^{-1}(\tau) = e^{i \tau \xi
}\hat \chi^l (x)\;\;,
$$
where the unitary transformation $\hat U (\tau ) = e^{i \tau \hat
Q }$ with $\tau$  a constant and $\xi $ an imaginary matrix.
\end{list}

\subsubsection*{GROUP XII: DISCRETE TRANSFORMATIONS: TIME REVERSAL}

\lista{Ax55}{A} \setcounter{Ax55}{55}
\item {\bf [S]} ${(\forall I_t)}_{\bf P_F} (I_t \den $ time
inversion, represented in the Hilbert space by the antiunitary
operator $\hat T \den {\tilde U} (I_t) )$.
\item {\bf [M]} \label{timerev} The field operators associated to
 basic fields transforms under time reversal as:
$$
\te {\hat \chi_{{\stackrel{\leftrightarrow}{AB}}_{12}}} (x)
\te^{-1}={D^{-1}(I_t)} {\hat \chi^*_{
{\stackrel{\leftrightarrow}{AB}}_{12}}} (I_t x) .
$$
\end{list}

\subsubsection*{GROUP XIII: LAGRANGE OPERATOR}

\lista{Ax57}{A} \setcounter{Ax57}{57}
\item {\bf [M]} \label{scalar} ${\hat {\cal L}} \equiv $
 differentiable Hermitian scalar-operator on any region $O \subset
 M^4$.
\item {\bf [M]} $(\forall \sg)_{\Sigma} (\exists {\hat {\cal L}})_{\cal A}
({\hat {\cal L}}[x] \equiv {\hat {\cal L}}[{\hat \chi}(x),
\partial_\mu {\hat \chi}(x)] )$. \footnote{We  shall use the
notation $\partial_\mu \den \frac{\partial}{\partial x^\mu}$.}
\item {\bf [S]} (${\hat {\cal L}}[x] \den $ local Lagrange
operator). In particular, (${\hat {\cal L}_{Kin}}[x] \den $ kinematic
Lagrange operator) and (${\hat {\cal L}_{Dyn}}[x] \den $ Dynamic
Lagrange operator).
\item {\bf [M]} (Generalized Momentum) \label{genemom} The generalized
momentum field operator ${{\hat \pi}^\mu}_l \defi \frac{\partial {\hat
{\cal L}}}{\partial (\partial_\mu {\hat \chi}^l)}$ will be constructed
with field operators as follows:
$$
{{\hat \pi}^\mu}_l \equiv {\hat \chi}^{r} ({\cal U}^\mu)_{r l}
$$ where the ${\cal U^\mu}$ are numerical matrices to be determined.
\item {\bf [M]} (Dynamical Lagrange Operator) \label{lagdin}
 The general form of the dynamical Lagrange operator is:
$$
\hat{\cal L}_{Dyn}= \sum_{n_1,n_2,...,n_s} a \cdot \hat \chi
\Pi_{k_1}^{n_1} ({\cal U}^\mu)^{k_1}\hat \chi \Pi_{k_2}^{n_2}
({\cal U}^\mu)^{k_2}\hat \chi \cdots \hat \chi \Pi_{k_s}^{n_s}
({\cal U}^\mu)^{k_s}\hat \chi.
$$
with the constraint that the sum $k_1+k_2+\cdots+k_s $ is even,
and where ``$a$" represents a real constant that must be different
for each member of the sum.
\end{list}

\subsubsection*{GROUP XIV: CONTINUOUS AND DISCRETE LAGRANGE TRANSFORMATIONS}
\lista{Ax62}{A} \setcounter{Ax62}{62}
\item {\bf [P]} \label{lagcon} (Invariance under transformations of the
 Poincar\'e group)
$$
{\hat U}(\Lambda,b) {\hat {\cal L}}[x] {\hat U}^{-1}(\Lambda,b)=
{\hat {\cal L}}[\Lambda x + b].
$$
\item {\bf [P]} (Invariance under gauge transformations)
$$
\hat U (\tau ) {\hat {\cal L}}[x] \hat U^{-1} (\tau ) = {\hat
{\cal L}}[x].
$$
\item {\bf [P]} \label{kinspacinv} (Kinematical invariance under space
inversion)
$$
\pa \hat {\cal L}_{Kin}[x] \pa^{-1} = \hat {\cal L}_{Kin} [I_s x].
$$
\item {\bf [P]} \label{kincharcon} (Kinematical invariance under charge
conjugation)
$$
\ca \hat {\cal L}_{Kin}[x] \ca^{-1} = \hat {\cal L}_{Kin} [x].
$$
\item {\bf [P]} \label{kintimerev}(Kinematical invariance under time
reversal)
$$ \te \hat {\cal L}_{Kin}[x] \te^{-1} = \hat {\cal
L}_{Kin}^* [I_t x].
$$
\end{list}

\subsubsection*{GROUP XV: STATIONARY ACTION PRINCIPLE}

\lista{Ax67}{A} \setcounter{Ax67}{67}
\item {\bf [P]} \label{stat}$(\forall \sg)_\Sg (\forall \hat {\cal L})_{\cal
 A}$. If ${\hat W}_{12} \defi
\int_{s_2}^{s_1} dx {\hat {\cal L}} [x]$ then
$$
\delta {\hat W}_{12}={\hat F}_1 - {\hat F}_2,
$$
where ${\hat F}_i \in {\cal A}$.
\item {\bf [S]} ${\hat W}_{12} \den$ Action integral operator.
\item {\bf [S]} $(\forall \sg)_\Sigma (\forall s_i)_{M^4} (\forall {\hat
F}_i)_{\cal A} ({\hat F}_i \rep $ the Hermitian generator of
infinitesimal  unitary transformation on the surface $s_i$).
\end{list}

{\footnotesize
{\bf Remark 1} As we can see from $\bf A_{45}$, $\bf A_{47}$ and $\bf
A_{54}$, the dimensionality of a given basic field will be determined
by the transformation properties under the group ${\bf L}_+$ enlarged
by space inversion $I_s$ (denoted by ${\bf L_C}$ and called complete
Lorentz group) and by $\hat Q$ (denoted by ${\bf L_{CQ}}$). For
instance, a scalar field transforms as the representation $(0,0)$ with
spin $j=0$ and it will be the only A-component of a field operator
representing a neutral basic field.  The next irreducible
representation is either $(0,\frac{1}{2})$ or $( \frac{1}{2},0 )$,
both corresponding to a field with spin $j= \frac{1}{2}$. But, in
order to represent the basic field of the electron, we have to
consider each of them as an A-component of the field operator i.e.,
the $(0,\frac{1}{2}) \oplus ( \frac{1}{2},0 )$ representation of the
Lorentz group enlarged by parity. This reflects the fact that the
Dirac representation used to describe the electron is
reducible. Therefore, an electro-positron basic field of spin
$\frac{1}{2}$ will be represented by an Hermitian field operator of
$8$ A-components (i.e. $2 \cdot 2 \cdot (2A+1)(2B+1)$).
{\bf Remark 2} Note that the requirement that the fields will be
characterized by Hermitian operators is not restrictive because we
work in a representation where  other properties are not
diagonalized.
{\bf Remark 3} A general field $\hat \chi$ representing a f-system
can be a very complicated mathematical entity as:
$$
\hat \chi =(\hat \chi_{(0,0)}, \hat \chi_{(\frac{1}{2},
\frac{1}{2})}, \hat
 \chi_{(0,\frac{1}{2})}, \hat
\chi_{(\frac{1}{2},0)}, \hat \chi_{(1,1)},\hat \chi_{(0,1)}, \hat
\chi_{(1,0)},
 \cdots)\;\;,
$$
 that can be written in a compact way as $\hat \chi=
 (\hat \phi, \hat \phi_\mu, \hat \nu, \cdots)$. In this case,
 the general field transforms as a reducible representation where
the matrix ${D(\La^{- 1})}$ includes the representations given by
the matrices ${\bf 1}, {\La^\mu}_\nu, $ 2-spinor, etc.
{\bf Remark 4} Since the general field operator contains all the
different representations, the $\bf A_{\ref{genemom}}$ states a
very natural assumption: that the generalized canonical momentum
field operators $\hat \pi^\mu_l$ are included in the components of
the general field $\hat \chi$. Note that the sum on the $r$-index
runs over all the components of the general field operator $\hat
\chi$. We use the following notation: when no confusion is
possible we shall consider the expression without matricial
indices, i.e., ${\hat \pi}^\mu=\hat \chi {\cal U}^\mu$.
{\bf
Remark 5} $\bf A_{\ref{genemom}}$ is not as restrictive as it
looks since the fields $\hat \chi$ include all the basic fields
represented of the f-system. Moreover, the properties of the
$({\cal U}^\mu)_{r l}$ matrices have not been specified yet. We
will see that these properties have very important consequences. }
%
%
%
\section*{3. SPIN-STATISTICS RELATION}
In this section we shall obtain the spin-statistic theorem. We shall present the
proofs of the theorems in an schematic way since our purpose is illustrating the 
role
of the axioms.

\lista{Th}{T}
\item ($\hat F (\delta \chi)$ Generating Operator) \label{gene} The
generating
 operator
 $\hat
F (\delta \chi)$ is given by:
$$
\hat F(\dchi) = \int ds \hat \pi_l \delta {\hat \chi}^l
$$

\begin{proof}
By steps: (1) The operator $ \hat F = \int_{s} ds_\mu ({{\hat
\pi}^\mu}_l {\tilde \delta {\hat \chi}^l}+  {\hat {\cal L}}\delta
x^\mu )$ is obtained from $\delta \hat W_{12}=\int_{s_2}^{s_1}
\delta (dx) {\hat {\cal L}}+  \int_{s_2}^{s_1} dx \delta {\hat
{\cal L}}$ and considering the expressions $\delta (dx) = dx
\partial_\mu \delta x^\mu$ and $\delta \hat {\cal L} =\tilde
\delta \hat {\cal L} + \frac{d \hat {\cal L}}{d x^\mu}\delta
x^\mu$ with $\tilde \delta \hat {\cal L}[x]= \frac{\partial {\hat
{\cal L}}}{\partial {\hat \chi}^l} {\tilde \delta \hat \chi^l }+
{{\hat \pi}^\mu}_l \partial_\mu (\tilde \delta \hat \chi^l)$ where
$\tilde \delta \hat \chi^l(x)$ is the total variation $\tilde
\delta \hat \chi^l(x)\defi \hat \chi'^l(x) - \hat \chi^l (x)$. (2)
Replacing ${\tilde \delta \hat \chi^l} = \delta \hat \chi^l -
\partial_\nu \hat \chi^l \delta x^\nu$ into the expression of
$\hat F$ we obtain $ \hat F(\dchi) \den \int_s ds_\mu  {{\hat
\pi}^\mu}_l \delta {\hat \chi}^l$ and $\hat F(\delta x) \den -
\int_s ds_\mu {{\hat T}^\mu}_\nu \delta x^\nu$ with ${{\hat
T}^\mu}_\nu \defi {{\hat \pi}^\mu }_l \partial_\nu {\hat \chi}^l -
{\hat {\cal L}} {\delta^\mu}_\nu.$ (3) Taking $ds_\mu = n_\mu ds$
and $\hat \pi_l \defi n_\mu {{\hat \pi}^\mu}_l$ (with $n_\mu \den$
a unit timelike vector and $ds \den$ the numerical measure of the
surface element) we have $ \hat F(\dchi) = \int ds \hat \pi_l
\delta {\hat \chi}^l $.
\end{proof}

\item (${\hat F}(\dpi)$ Generating Operator) \label{fpi} The
 generating operator ${\hat F} (\dpi)$ is given by:
$$
{\hat F'} = {\hat F}(\dpi) =-\int ds \delta {\hat \pi}_l {\hat
\chi}^l\;\;.
$$

\begin{proof}
(1) Consider the Lagrangian ${\hat {\cal L'}} = {\hat {\cal
L}}-\partial_\mu \hat f^\mu$ (which is equivalent to ${\hat {\cal
L}}$), from ${\bf A_{68}}$ we have $\delta \hat {W'}_{12} =\hat
{F'}_1 - \hat {F'}_2 $ with $\hat {F'}_i={\hat F}_i-\delta {\hat
W}_i$ and where ${\hat W}_i \defi \int_{s_i} d{s_\mu} \hat f^\mu$.
(2) Taking ${\hat f}^\mu = {{\hat \pi}^\mu}_l {\hat \chi}^l $ into
the expression of the generator ${\hat F'}$ and using $\bf
T_{\ref{gene}}$.
\end{proof}

\item (Symmetric Infinitesimal Generator) \label{hercon} If $\hat F^{sym} \defi
\frac{1}{2}[{\hat F}(\dchi) + {\hat F}(\dpi) ]$ then
$$
\hat F^{sym} = \frac{1}{2} \int ds [{\hat \chi}^{r} ({\cal
U}^\mu)_{r l} \delta \hat \chi^l - \delta \hat \chi^{r} ({\cal
U}^\mu)_{r l} \hat \chi^l]\;\;\;\;\;\;\mbox{where}\;\;\;\;\; {\cal
U}^{\dag \mu} =-{\cal U}^\mu\;\;.
$$

\begin{proof}
Using $\bf T_{\ref{gene}}$, $\bf T_{\ref{fpi}}$, $\bf
A_{\ref{hermit}}$, $\bf A_{\ref{genemom}}$ and the hermiticity
condition of $\hat F$ stated in $\bf A_{\ref{stat}}$.
\end{proof}

\item (Equivalent Generating Operators) \label{genequi}
(a) The generating
 operators
${\hat F}(\dchi)$, ${\hat F}(\dpi)$ and $\hat F^{sym}$ are
equivalents.
(b) $ {\hat \pi}_l \delta {\hat \chi}^l = -
\delta{\hat \pi}_l {\hat \chi}^l\;\;.$

\begin{proof}
They are obtained from equivalents Lagrangian (i.e. differing each
other by a divergence). Then, use the expressions of ${\hat
F}(\dchi)$ and ${\hat F}(\dpi)$.
\end{proof}

\item \label{descom} The matrices ${\cal U}^\mu$ can be decomposed
in a symmetric and an antisymmetric parts, that is, ${\cal
U}^\mu={\cal U}^\mu_S + {\cal U}^\mu_A$ where each part satisfy:
$$
{{\cal U}^\mu_S}^T = {\cal U}^\mu_S,\;\;\;\;\;
{{\cal U}^\mu_A}^T = - {\cal U}^\mu_A.
$$

\begin{proof}
From matrix algebra.
\end{proof}

\item \label{realid} The symmetric matrices ${\cal U}^\mu_S$ are
imaginary and the antisymmetric matrices ${\cal U}^\mu_A$ are
real, i.e.:
$$
({\cal U}^\mu_S)^* = -{\cal U}^\mu_S,\;\;\;\;\;
({\cal U}^\mu_A)^* = {\cal U}^\mu_A.
$$

\begin{proof}
From $\bf T_{\ref{hercon}}$ and $\bf T_{\ref{descom}}$.
\end{proof}

\item \label{twofield} There are two different classes of field
operators that satisfy the following commutation relations
according to the different properties of the matrices ${\cal
U}^\mu$:
$$
[{\cal U}^\mu_S {\hat \chi}(x),\delta {\hat
\chi}(x')]_{+}=0\;\;\;\;\;,\;\;\;\;\; [{\cal U}^\mu_A {\hat
\chi}(x),\delta {\hat \chi}(x')]_{-}=0.
$$

\begin{proof}
\begin{enumerate}
\item Let ${\cal U}^\mu$ be symmetric, then ${\hat \chi} {\cal
U}^\mu_S = {\cal U}^\mu_S {\hat \chi}$. Thus we have from $\bf
T_{\ref{genequi}}$:
$$
({\cal U}^\mu_S {\hat \chi}) \delta{\hat \chi} = - \delta {\hat
\chi} ({\cal U}^\mu_S {\hat \chi}) \Rightarrow [({\cal U}^\mu_S {\hat
\chi}),\delta{\hat \chi}]_+ = 0.
$$
\item Let ${\cal U}^\mu$ be antisymmetric, then ${\hat \chi} {\cal
U}^\mu_A = - {\cal U}^\mu_A {\hat \chi}$. So, again from $\bf
T_{\ref{genequi}}$:
$$
(-{\cal U}^\mu_A {\hat \chi}) \delta{\hat \chi} = - \delta {\hat
\chi} ({\cal U}^\mu_A {\hat \chi}) \Rightarrow [({\cal U}^\mu_A {\hat
\chi}), \delta{\hat \chi}]_- = 0.
$$

These commutation relations have been obtained for only one point
$x$. The expressions for arbitrary different points $x$ and $x'$ are
obtained by the compatibility requirement for operators located at
distinct points of a spacelike surface.
\end{enumerate}
\end{proof}
\end{list}

\lista{Df5}{D} \setcounter{Df5}{5}
\item (Bose and Fermi Field Operators) \label{bosfer}
\item [(a) Fermi Operators:] The field operators that satisfy the
first group of commutation relations of $\bf T_{\ref{twofield}}$ will
be called {\em Fermi field operators} and will be denoted by $\hat
\psi$.
\item [(b) Bose Operators:] The field operators that satisfy the
second group of the commutation relations of $\bf T_{\ref{twofield}}$
will be called {\em Bose field operators} and will be denoted by $\hat
\phi$.
\end{list}

{\footnotesize {\bf Remark} $\bf D_{\ref{bosfer}}$ conventionally
assigns a name to field operators associated to symmetric and
antisymmetric matrices ${\cal U}^\mu$ that satisfy the
anticommutation or commutation relations respectively. Thus, it is
clear that we have not obtained the spin-statistic relation,
because we have not specified any spin value for each different
class of field operators.}

\lista{Th7}{T} \setcounter{Th7}{7}
\item (Lagrange Operator) \label{lagform} The Lagrange operator
can be expressed in the general form:
$$
{\hat {\cal L}}=\hat {\cal L}_{Kin}+\hat {\cal L}_{Dyn}
$$
where
$$
\hat {\cal L}_{Kin} \den \frac{1}{2}[ {\hat \chi}^{r} ({\cal
U}^\mu)_{r l} \partial_\mu {\hat \chi}^l -  \partial_\mu {\hat
\chi}^{r} ({\cal U}^\mu)_{r l} {\hat \chi}^{l} ]
,\;\;\;\;\; \hat {\cal L}_{Dyn} \den - {{\hat
T}^\mu}_\mu.
$$

\begin{proof} From the expression of ${{\hat T}^\mu}_\nu = {{\hat \pi}^\mu }_l
\partial_\nu {\hat \chi}^l - {\hat {\cal L}} {\delta^\mu}_\nu$ (obtained from $\bf
A_{\ref{stat}}$) with $\mu=\nu$, and using $\bf A_{\ref{genemom}}$, $\bf
A_{\ref{hermit}}$, and $\bf T_{\ref{hercon}}$.  But, in order to have an Hermitian
operator ($\bf A_{\ref{scalar}}$) we must add a term, i.e.,
$$
{\hat {\cal L}}= {\hat \chi}^{r} ( {\cal U}^\mu)_{r l}
\partial_\mu {\hat \chi}^l - {{\hat T}^\mu}_\mu - \frac{1}{2}
\partial_\mu [ \hat \chi^{r}({\cal U}^\mu)_{r l} \hat \chi^l
].
$$
\end{proof}

\item \label{reptime} The field operator associated to an $f$-system
 transforms under temporal inversion as a reducible representation,
 i.e.,
$$
\te \hat \chi (x) \te^{-1} = {\bf D}^{-1}(I_t) \hat \chi^* (I_t x)
$$
with
$$ {\bf D}(I_t) = \pmatrix{D_1 & 0& \cdots & 0 \cr 0 & D_2 & \cdots &
0 \cr \vdots & \vdots & \ddots & 0 \cr 0 & 0 & 0 & D_j }
$$ where $D_j (I_t) \den$ a matrix transformation of the $j$-th basic
field.

\begin{proof}
From $\bf A_{\ref{field}}$ and $\bf A_{\ref{timerev}}$.
\end{proof}

\item (T Transformation) \label{tramatr} The matrices ${\cal U}^\mu $
transform under time reversal as:
$$
{\bf D}(I_t){\cal U}^\mu {\bf D}^{-1}(I_t) = (-1)^{\delta_{\mu 0}} (\pm) {\cal
 U}^\mu
$$ where the matrices ${\bf D}(I_t)$ are imaginary for Fermi field
operators and real for Bose field operators, with $(+)$ for
antisymmetrical and $(-)$ for symmetrical ${\cal U}^\mu$ matrices.

\begin{proof}
We split the general field operator $\hat \chi$ in terms of Bose and
 Fermi field operators, i.e. $\hat \chi= (\hat \phi, \hat
 \psi)$. According to $\bf T_{\ref{twofield}}$, the kinematical
 Lagrange operator expressed in this way (it is sufficient to consider
 the asymmetrical version) is,
$$ \hat {\cal L}_{Kin}= \hat \chi {\cal U}^\mu \partial_\mu \hat \chi
= \hat \phi {\cal U}^\mu_A \partial_\mu \hat \phi + \hat \psi {\cal
U}^\mu_S \partial_\mu \hat \psi,
$$
applying the time reversal operator $\bf A_{\ref{timerev}}$ we have:
$$ \te \hat{\cal L}_{Kin} \te^{-1} = \hat \phi^* {\bf D} {\cal
U}^\mu_A {\bf D}^{-1} \overline{\partial_\mu} \hat \phi^* + \hat
\psi^* {\bf D} {\cal U}^\mu_S {\bf D}^{-1} \overline{ \partial_\mu}
\hat \psi^* ,
$$
where
$$
\overline{ \partial_\mu}= (-1)^{\delta_{\mu 0}} \partial_\mu ,
$$
Comparing with (see $\bf A_{\ref{kintimerev}}$),
$$
\hat {\cal L}_{Kin}^*= \hat \phi^* {{\cal U}^\mu_A}^* \partial_\mu
\hat \phi^* + \hat \psi^* {{\cal U}^\mu_S}^* \partial_\mu \hat \psi^*
,
$$
we finally obtain, from $\bf T_{\ref{realid}}$:
$$
{\bf D} {\cal U}^\mu_A {\bf D}^{-1} = (-1)^{\delta_{\mu 0}} {\cal
U}^\mu_A,
$$
$$
{\bf D} {\cal U}^\mu_S {\bf D}^{-1} = (-1)^{\delta_{\mu 0}} (-) {\cal
 U}^\mu_S.
$$
\end{proof}
\end{list}
\mbox{}

Independently of the last theorem, we can prove a property of the
${\bf D}(I_t)$ matrix:
\lista{Th10}{T} \setcounter{Th10}{10}
\item \label{dimag} The matricial representation ${\bf D}(I_t)$ is
imaginary only for fields of half integer spin and real for fields
of integer spin.

\begin{proof}
We must consider complex Lorentz transformations acting upon an
 A-component of $\hat \chi$.  Thus, we define $\hat J_{i 4}\defi i
 \hat J_{i 0}$ with $\hat J_{i0}^* = - \hat J_{i0}$ so that $\hat
 J_{i4}^* = \hat J_{i4}$.  The $\hat T$ operator of $\bf
 A_{\ref{timerev}}$ can be written as a rotation of $\hat P_i$ where
 $\hat P_i$ is the $x_i$ space-inversion operator,
$$
\hat T= e^{-i\frac{\pi}{2}\hat J_{i4}} \hat P_i
e^{+i\frac{\pi}{2}\hat J_{i4}}
$$
using that $\hat P_i \hat J_{i0}\hat P_i^{-1}=-\hat J_{i0}$ we have,
$$
\hat T= e^{-i\pi \hat J_{i4}} \hat P_i,
$$
complex conjugating and replacing,
$$
\hat T^*= e^{+i\pi \hat J_{i4}} \hat P_i=e^{+2i\pi \hat J_{i4}}
\hat T.
$$

Since in the $(A,B)$ representation of Lorentz group, the operator $i
 \hat J_{i0}$ is given, in our new notation, by $\hat J_{i4}= \hat A-
 \hat B$. Denoting the eigenvalues of $\hat J_{i4}$ by $j$, we have
 for the matrix representation of $\hat T$:
$$
{D(I_t)^*}= e^{+2i\pi j} D(I_t)
$$
with $D(I_t)^*=D(I_t)$ for $j$ integer and $D(I_t)^*= -D(I_t)$ for
$j$ half integer. The theorem follows since the matrix ${\bf
D}(I_t)$ is a reducible representation as given by $\bf
T_{\ref{reptime}}$.
\end{proof}
\end{list}
\mbox{}

We will close this section with a fundamental result of any
relativistic quantum theory of fields:
\lista{Th11}{T} \setcounter{Th11}{11}
\item (Spin-Statistics Relation) Fields with half integer spin are
represented by Fermi field operators, and fields with integer spin
are represented by Bose field operators.

\begin{proof}
Comparing $\bf T_{\ref{dimag}}$ with $\bf T_{\ref{tramatr}}$,
noting that they are obtained independently.
\end{proof}
\end{list}

{\footnotesize {\bf Remark 1}
The derivation of the spin-statistics theorem presented here follows the proof
presented by
Schwinger in ref.\cite{Sc51,Sc53}. Modifications of his original derivation based
on
different assumptions were presented by himself on several occasions during the
rest of his
life (see for example \cite{Sc70}). A nice and comprehensive account of this
theorem can be
found in ref.\cite{Su97}.
{\bf Remark 2} The explicit use of the so called ``local causality"
requirement, expressed as the commutation relation between field
operators, is not necessary in this approach as it is the case for the
proof of free fields (see ref. \cite{Pa40, We64}) or for interacting
fields (see ref. \cite{Lu58, Bu58}). Thus, it can seem that this
assumption is not needed in our proof. However, as we will see,
``local causality" is a consequence derivable from the expression of
the generating operators. }
%
%
%
\section*{4. COMMUTATION RELATIONS}
In this section the commutation relations of field operators will be
deduced from a very general property of the generators. First we state
a theorem concerning this general property,
\lista{Th12}{T} \setcounter{Th12}{12}
\item \label{generu} Let $\hat F$ be an infinitesimal Hermitian
 generator. The unitary operator defined by ${\hat U} \defi 1 +i {\hat
 F}$ acts on an arbitrary operator $({\hat A})_{\cal A}$ as:
$$
\delta {\hat A}= \frac{1}{i}[{\hat A},{\hat F}] .
$$

\begin{proof}
Evaluate ${\hat A}'={\hat U} {\hat A} {\hat U}^{-1}$.
\end{proof}

\item \label{comgen} The commutation relations of the field
operators with the generators are given by:
$$
[{\cal U}^\mu \hat \chi (x) , \hat F(\delta \chi)]_{\pm}=i {\cal
U}^\mu \delta \hat \chi.
$$

\begin{proof}
Using $\bf T_{\ref{generu}}$.
\end{proof}

\item (``Equal-time" Commutation Relations) \label{comfield} The
covariant generalization of the equal-time commutation relations
of the field operators are given by:
$$
[{\cal U}^\mu \hat \chi (x) , \hat \chi (x'){\cal U}^\mu ]_\pm = i {\cal U}^\mu
 \delta_s
 (x-x')
$$ where $\delta_s (x-x')$ is defined by $f(x) \defi \int_s ds'
\delta_s (x-x') f(x')$. The commutators $(- )$ are used for Bose field
operators and anticommutators $(+)$ for Fermi field operators.

\begin{proof}
From $\bf T_{\ref{comgen}}$ and using the expression of $\hat F
(\delta \chi)$ (or its symmetric version).
\end{proof}

\item (Bose and Fermi ``Equal-time" Commutation Relations) In terms
of Bose and Fermi field operators, we have:
$$ [{\cal U}^\mu_A \hat \phi (x) , \hat \phi (x') {\cal U}^\mu_A ]_- =
i {\cal U}^\mu_A \delta_s (x-x') ,
$$
$$ [{\cal U}^\mu_S \hat \psi (x) , \hat \psi (x') {\cal U}^\mu_S ]_+ =
i {\cal U}^\mu_S \delta_s (x-x') ,
$$
$$ [{\cal U}^\mu_S \hat \psi (x) , \hat \phi (x') {\cal U}^\mu_A ]_\pm
= 0.
$$

\begin{proof}
From $\bf T_{\ref{comfield}}$ and $\bf D_{\ref{bosfer}}$.
\end{proof}
\end{list}

{\footnotesize
{\bf Remark 1} $\bf T_{\ref{comgen}}$ can equally
be proved using the symmetric expression $\hat F^{sym}$ given in
$\bf T_{\ref{hercon}}$.
{\bf Remark 2} The commutation relations
are a covariant generalization of the commutation relations for
equal-time. To see this, note that for $n_\mu = (1,0,0,0)$ we have
$ds=d^3x$ and $\delta_s(x-x')$ become $\delta^3(x-x')$.
{\bf
Remark 3} Since the above fields are characterized by Hermitian
operators their commutators does not vanish. However, it is clear
that in terms of the non-hermitian operators like $\hat \chi^+$
and $\hat \chi^-$, they vanish. }
%
%
%
\section*{5. CPT THEOREM}
In this section we give another fundamental theorem that must be
 satisfied by any relativistic quantum theory of fields. In order to
 deduce it, we must obtain first some preliminary results,

\lista{Th16}{T} \setcounter{Th16}{16}
\item \label{prep} A matrix
representation of $\hat P$ is given by,
$$
D^{-1}(I_s) = \nu_s \pmatrix{0&1\cr 1& 0}
\;\;\;\;\mbox{for}\;\;\;\;A \ne B
$$
$$
D^{-1}(I_s) = \nu_s \;\;\;\;\mbox{for}\;\;\;\;A = B.
$$

\begin{proof}
From $\bf A_{\ref{spacinv}}$.
\end{proof}

\item \label{pprep} The matrix representation of $\hat P$ acting
upon a field operator ${\hat
\chi_{{\stackrel{\leftrightarrow}{AB}}_{12}}}$ to
${\stackrel{\leftrightarrow}{AB_1}} \ne
{\stackrel{\leftrightarrow}{AB_2}}$ is given by:
$$
D^{-1}(I_s) =\pmatrix{ D^{-1}_1 (I_s) & 0 \cr 0 & D^{-1}_2 (I_s) }.
$$

\begin{proof}
From $\bf T_{\ref{prep}}$ and $\bf A_{\ref{spacinv}}$.
\end{proof}

\item \label{redspa} A field operator $\hat \chi$ associated to an
f-system transforms under spatial inversion as a reducible
representation, i.e.,
$$
\pa \hat \chi (x) \pa^{-1} = {\bf D}^{-1}(I_s) \hat \chi (I_s x)
$$
with
$$ {\bf D}(I_s) = \pmatrix{D_1 & 0& \cdots & 0 \cr 0 & D_2 & \cdots &
0 \cr \vdots & \vdots & \ddots & 0 \cr 0 & 0 & 0 & D_j }
$$ where $D_j (I_s) \den$ a matrix transformation for the $j$-th basic
field.

\begin{proof}
From $\bf A_{\ref{field}}$, $\bf A_{\ref{spacinv}}$ and $\bf
T_{\ref{pprep}}$.
\end{proof}

\item \label{redchar} A field operator $\hat \chi$ associated to an
f-system transforms under charge conjugation as a reducible
representation, i.e.,
$$
\ca \hat \chi (x) \ca^{-1} = {\bf D}^{-1}(I_c) \hat \chi (I_c x)
$$
with
$$
{\bf D}(I_c) =\pmatrix{D_1 & 0& \cdots & 0 \cr 0 & D_2 & \cdots & 0
\cr \vdots & \vdots & \ddots & 0 \cr 0 & 0 & 0 & D_j }
$$
where $D_j (I_c) \den$ a matrix transformation of the $j$-th basic field.

\begin{proof}
From $\bf A_{\ref{field}}$ and $\bf A_{\ref{charcon}}$.
\end{proof}

\item (P Transformation) \label{dese} The matrices ${\cal U}^\mu$
transform under space inversion as:
$$
{\bf D}(I_s){\cal U}^\mu {\bf D}^{-1}(I_s) =- (-1)^{\delta_{\mu 0}}
{\cal U}^\mu.
$$

\begin{proof}
From $\bf A_{\ref{spacinv}}$,  $\bf T_{\ref{redspa}}$ and $\bf
T_{\ref{lagform}}$.
\end{proof}

\item (PT Transformation) \label{pt} The ${\cal U}^\mu$ matrices
transform under the combined $PT$ transformation as:
$$
{\bf D}(I_s) {\bf D}(I_t){\cal U}^\mu {\bf D}^{-1}(I_t) {\bf D}^{-1}(I_s) = \mp
 {\cal U}^\mu
$$
where $(-)$ and $(+)$ corresponds to the antisymmetric and symmetric matrices
 respectively.

\begin{proof}
From $\bf T_{\ref{tramatr}}$ and $\bf T_{\ref{dese}}$.
\end{proof}

\item (C Transformation) \label{co} The matrices ${\cal U}^\mu$
transform under charge conjugation as:
$$
{\bf D}(I_c) {\cal U}^\mu {\bf D}^{-1}(I_c)= {\cal U}^\mu.
$$

\begin{proof}
From $\bf A_{\ref{charcon}}$,  $\bf T_{\ref{redchar}}$ and $\bf
T_{\ref{lagform}}$.
\end{proof}

\item (CPT Transformation) \label{cpt} The Lagrange operator is
invariant under the antiunitary $\ca \pa \te$ transformation:
$$
\ca \pa \te \hat {\cal L}[x] (\ca \pa \te)^{-1} = \hat {\cal L}^* [-x] .
$$

\begin{proof}
The invariance of the kinematical part $\hat {\cal L}_{Kin}$ is
trivially proved from the requirements of $\bf
A_{\ref{kinspacinv}}$, $\bf A_{\ref{kincharcon}}$ and $\bf
A_{\ref{kintimerev}}$. That of the dynamical part follows taking
into account that $\hat{\cal L}_{Dyn}$ is expressed as an even
combination of ${\cal U}^\mu$ matrices, i.e. from $\bf
A_{\ref{lagdin}}$, and using $\bf T_{\ref{pt}}$ and $\bf
T_{\ref{co}}$ we have:
$$ \ca \pa \te \hat{\cal L}_{Dyn} [x] (\ca \pa \te )^{-1}=
\sum_{n_1,n_2,...,n_s} a \cdot \hat \chi^* (-x) \Pi_{k_1}^{n_1}
(\mp{\cal U}^\mu)^{k_1}\hat \chi^* (-x) \times
$$
$$ \Pi_{k_2}^{n_2} (\mp{\cal U}^\mu)^{k_2}\hat \chi^* (-x) \cdots \hat
 \chi^* (-x) \Pi_{k_s}^{n_s} (\mp{\cal U}^\mu)^{k_s}\hat \chi^* (-x)
 ,
$$
taking the complex conjugate (by the antiunitary requirement of $\te$),
$$ \{ \ca \pa \te \hat{\cal L}_{Dyn}[x] (\ca \pa \te )^{-1} \}^* =
\sum_{n_1,n_2,...,n_s} a \cdot \hat \chi (-x) \Pi_{k_1}^{n_1}
{(\mp{{\cal U}^\mu}^*)}^{k_1} \hat \chi(-x) \times
$$
$$ \Pi_{k_2}^{n_2} {(\mp{{\cal U}^\mu}^*)}^{k_2}\hat \chi (-x) \cdots
\hat \chi (-x) \Pi_{k_s}^{n_s} {(\mp{{\cal U}^\mu}^*)}^{k_s}\hat
\chi(-x) ,
$$
since the ${\cal U}^\mu$ matrices are real and imaginary to the
parts antisymmetrical and symmetrical, we have:
$$
= \sum_{n_1,n_2,...,n_s}a \cdot \hat \chi (-x) \Pi_{k_1}^{n_1}
{(-{{\cal U}^\mu})}^{k_1}\hat \chi (-x) \Pi_{k_2}^{n_2} {(-{{\cal
U}^\mu})}^{k_2} \hat \chi (-x) \cdots
$$
$$ \cdots \hat \chi (-x) \Pi_{k_s}^{n_s} {(-{{\cal U}^\mu})}^{k_s}\hat
\chi(-x) ,
$$
with the restriction of $\bf A_{\ref{lagdin}}$ that the sum
$k_1+k_2+\cdots+k_s $ is even, follows:
$$
\{ \ca \pa \te  \hat{\cal L}_{Dyn} [x] (\ca \pa \te )^{-1}\}^* = \hat{\cal
 L}_{Dyn}[-x].
$$
\end{proof}
\end{list}

{\footnotesize {\bf Remark 1} $\bf T_{\ref{cpt}}$ implies that the
antiunitary operator $\ca \pa\te$ describes a symmetry
transformation for any local relativistic quantum field theory.
Note that thanks to our abstract formulation, the proof follows
without to state the transformation properties for each particular
representation of the field operators.}
%
%
%
\section*{6. EXAMPLE: SPIN $1$ AND SPIN $\frac{1}{2}$ FIELDS IN
  INTERACTION }

In this section we present an example, that follows from our general
 formalism in the case of a system of interacting fields of spin $1$
 and spin $\frac{1}{2}$. In this particular framework, the Lagrange
 operator and the expression for the generators will be obtained.
 First, an useful definition:

\lista{Df7}{D} \setcounter{Df7}{7}
\item (Fundamental and Non-fundamental Field Operators) \label{fundam}
 Let the timelike vector $n_\mu$ \linebreak $=(1,0,0,0)$ be such that
 $ds_\mu \equiv dx^3$. The field operators $\hat \pi^0_l$ and $\hat
 \chi^l$ that appear in the expressions of the generators will be
 called {\em fundamental field operators} and they are the independent
 variables which obey the equations of motion. The remainder ones will
 be called {\em non-fundamental field operators} and they are the
 dependent variables which obey the constraint equations.
\end{list}

Now, we consider the general field:
$$
\hat \chi = (\hat \chi_{(\frac{1}{2},\frac{1}{2})}, \hat \chi_{(0,1) \oplus
 (1,0)}, \hat
 \chi^1_{{ (0,
\frac{1}{2}) \oplus (\frac{1}{2},0) }}, \hat \chi^2_{{ (0, \frac{1}{2}) \oplus
 (\frac{1}{2},0) }}  )
\equiv ( \hat A_\nu, \hat F_{\mu \nu},\hat \psi_1, \hat \psi_2 ) \den (\hat
 \phi, \hat
 \psi )
$$
where $\hat \phi$ and $\hat \psi$ have $10$ and $8$ A-components respectively
 with,
$$
\hat \phi \den (\hat A_\nu, \hat F_{\mu \nu} ) \;\;\;\;\mbox{ and  }\;\;\;\;
\hat \psi \den (\hat \psi_1, \hat \psi_2 ) \;\;,
$$
describing basic fields of $1$ and $\frac{1}{2}$ spin.
With the matrices,
$$
{\cal U}^\mu_A= \pmatrix{0 & 1\cr -1 & 0}
\;\;\;\;\;\;\;\mbox{ and }\;\;\;\;\;\;\;
{\cal U}^\mu_S=i \alpha^\mu {\bf 1}= \pmatrix{i \alpha^\mu & 0 \cr 0 & i
 \alpha^\mu},
$$
the kinematical Lagrangian can be written as (using $\bf
T_{\ref{lagform}}$):
$$
\hat {\cal L}_{Kin}= \hat {\cal L}_{Kin}^{1} + \hat {\cal
 L}_{Kin}^{\frac{1}{2}},
$$
with:
$$
\hat {\cal L}_{Kin}^{1}= - \hat F^{\mu \nu} \partial_\mu \hat A_\nu + \hat A_\nu
\partial_\mu \hat
F^{\mu \nu}
\;\;\;\;\mbox{ and }\;\;\;\;
\hat {\cal L}_{Kin}^{\frac{1}{2}}= i \hat \psi \alpha^\mu {\bf 1}\partial_\mu
 \hat
 \psi.
$$
The dynamical Lagrangian assume the following form (using $\bf
A_{\ref{lagdin}}$):
$$ \hat {\cal L}_{Dyn}= \hat {\cal L}_{Dyn}^{1} + \hat {\cal
L}_{Dyn}^{\frac{1}{2}} + \hat {\cal L}_{Dyn}^{int},
$$
where the interaction terms of the fields with themselves (i.e.
``autointeraction") are given by:
$$
\hat {\cal L}_{Dyn}^{1}= \hat \phi {\cal U}^\mu_A {\cal U}^\mu_A
\hat \phi = - a \hat A_\mu \hat A^\mu - b \hat F_{\mu \nu} \hat F^{\mu
\nu},
$$
$$
\hat {\cal L}_{Dyn}^{\frac{1}{2}} = \hat \psi {\cal U}^\mu_S {\cal
U}^\mu_S \hat \psi = - c \hat \psi \alpha^\mu {\bf 1} \alpha^\mu \hat
\psi.
$$
The interaction Lagrangian (between different fields) must be
conjectured, but taking into account the transformation properties of
the Lagrange operator, we propose:
$$
\hat {\cal L}_{Dyn}^{int}= \hat A_\mu {\cal U}^\mu_A {\cal U}^\mu_A  \hat \psi
 {\cal
 U}^\mu_S
{\cal U'}^\mu_A \hat \psi\;\;\;\;\mbox{ with }\;\;\;\;{\cal U'}^\mu_A \equiv
 \xi.
$$
Evaluating the generators we have:
$$
\hat F(\delta A_\nu, \delta F^{\mu \nu}, \delta \psi ) = - \int ds_0 (\hat F^{0
 \nu}
 \delta \hat A_\nu - i
\alpha^0 \hat \psi \delta \hat \psi ) .
$$
It follows that the fundamental fields are (using $\bf
D_{\ref{fundam}}$) $(\hat A_k, \hat F^{0 k}, \hat \psi)$ and the
non-fundamental fields are $(\hat A_0, \hat F^{k l})$. Obviously,
we do not pretend to obtain electrodynamics, since we have only
demanded gauge invariance of the first kind.
%
%
%
\section*{7. CONCLUSIONS}
\label{Discussion} We have presented a physical axiomatization of
the RQFT which has a number of important advantages:

\begin{enumerate}

\item The requirement of kinematical invariance under time reversal
imposes a restriction upon the field operators: the
spin-statistics relation. Pauli's proof \cite{Pa40} of
spin-statistics proceeds differently for integer and half-integer
spin. The proofs given by L\"uders and Zumino \cite{Lu58} and,
similarly, by Burgoyne \cite{Bu58} for interacting fields, follow
an indirect argument: the wrong relation between spin and
statistics cannot be postulated in a relativistic quantum field
theory without to come in contradiction. On the other hand,
Weinberg \cite{We64} shows that ``causality" is satisfied only
with the correct statistics and with crossing symmetry. In all
these cases the theorem is proved invoking the ``causality"
requirement. Here we follow a direct argument for both integer and
half-integer spin and, in contrast with the other formulations,
without the explicit use of the ``causality" requirement.

\item The commutation relations of the field operators are not
 postulated as usual: they are derived from the stationary action
 principle, in particular from the generators associated with a given
 spacelike surface.  However, the condition of physical independence
 of different points of a spacelike surface, is implicitly expressed
 in the structure of the generating operators because these are
 constructed from field operators attached on a spacelike surface.

\item The CPT theorem is proved using the mathematical advantages of
this formulation: the different transformation properties of the
${\cal U}^\mu$ matrices and the general form postulated of the
dynamical Lagrangian. As it is well known, the proof given by
L\"uders \cite{Lu57} of CPT theorem uses the spin-statistics
relation. In our case this assumption is stated in the different
transformation properties of the two kind of ${\cal U}^\mu$
matrices which are related to each statistics. Moreover, our proof
is general because it follows without to study the different
transformation properties of each particular field.
\end{enumerate}

We must point out that the quantum theory of fields is a very general
 framework that, by adding special postulates and subsidiary
 assumptions, reduces to particular theories as the known case of
 electrodynamics. In the last section we provide an example of this
 reduction for an interacting f-system.

The proof of the spin-statistics relation is presented here into the relativistic
framework. However, a recent paper by Peshkin \cite{Peshkin03} has tried to prove 
the
spin-statistics theorem from rotational properties of the non-relativistic wave
fuction for the case of spinless particles opening thus a vivid debate on the 
subject.

However, that proof is based on the assumption that exchange of
identical partical can be represented as a physical
transportation. This is a misundestanding, as has been pointed out in
several references \cite{Allen03,Sudarshan03}, that the use of a
formal semantic would have helped to avoid. Indeed, as Sudharhan
\cite{Sudarshan03} has stated  that this kind of problems
``would not arise if we were to exercise perfect semantic precision".

Shaji and Sudharsan \cite{Shaji03} have developed another proof, much
more clear and physically well motivated. It hinges, however, on the
posibility of representing non-relativistic fields as hermitean
operators. In our previous axiomatization, we concluded that Galilean
invariance forbids such a representation. Indeed, no Galilean field
operator of non-null mass can be hermitian (see \cite{JMLL67,Pu01})
because Galilean group admits only projective representations. The
proof may be valid for a ``gas in a box'' and similar systems, which
are models of macroscopic bodies, since Galilean invariance is broken
for those systems. However, this issue merits a separate paper for its
analysis.

Last but not least, in this approach fields are considered as {\em
things with properties} which are represented by operators that
satisfy certain symmetry transformations. This means that, in our
conception, fields are more fundamental than both: particles and
symmetries, since symmetries are symmetries of properties of
things and without things, there are no symmetries. Indeed, fields
are unobservable but they should not be regarded as auxiliary
devices with no physical meaning since the concept of quantum
fields provides a mechanism of interaction explicitly expressed by
the form of the Lagrangian operator. On the contrary, this
mechanism is lost if fields and their mathematical referents are
regarded only as auxiliary computational devices. In sum, we think
that the concept of field must be regarded as a deep basic
hypothesis in term of which we try to explain the behavior of
matter.
%
%
%
\section*{ACKNOWLEDGMENTS}
The authors would like to thank S. Perez Bergliaffa for
helpful comments and M. A. Bunge for a critical reading of the
manuscript. HV is member of CONICET and
acknowledges support of the University of La Plata. GP acknowledges support from
 FOMEC scholarship program.
%
%
%

\end{document}